# Interfacial Cooling and Heating, Temperature Discontinuity and Inversion in Evaporation and Condensation

Gang Chen

Department of Mechanical Engineering
Massachusetts Institute of Technology

**Abstract**
Although ubiquitous in nature and industrial processes, transport processes at the interface during evaporation and condensation are still poorly understood. Experiments have shown temperature discontinuities at the interface during evaporation and condensation but the experimentally reported interface temperature jump varies by two orders of magnitude. Even the direction of such temperature jump is still being debated. Using kinetic-theory based expressions for the interfacial mass flux and heat flux, we solve coupled problem between the liquid and the vapor phase during evaporation and condensation. Our model shows that when evaporation or condensation happens, an intrinsic temperature difference develops across the interface, due to the mismatch of the enthalpy carried by vapor at the interface and the bulk region. The vapor temperature near the interface cools below the saturation temperature on the liquid surface during evaporation and heats up above the latter during condensation. However, many existing experiments have shown an opposite trend to this prediction. We explain this difference as arising from the reverse heat conduction in the vapor phase. Our model results compare favorably with experiments on both evaporation and condensation. We show that when the liquid layer is very thin, most of the applied temperature difference between the solid wall and the vapor phase happens at the liquid-vapor interface, leading to saturation of the evaporation and the condensation rates and the corresponding heat transfer rate. This result contradicts current belief that the evaporation and condensation rates are inversely proportional to the liquid film thickness.

## I. INTRODUCTION

Evaporation and condensation processes happen widely in nature and industrial technologies [1–4]. Yet, our current understanding of interfacial transport during evaporation and condensation is still limited. For example, interfacial temperature discontinuities have been experimentally observed [5–8]. Kinetic theory-based modeling [9–13], Monte Carlo simulations [14], and molecular dynamics simulations[15–19] have also predicted the existence of this temperature discontinuity. However, experimentally measured temperature discontinuities vary widely, and the kinetic theory so-far predicts a temperature discontinuity that is usually 1-2 orders of magnitude smaller than some experimental observations [2,8]. There are conflicting views on even the signs of the temperature discontinuities. For example, most experiments on evaporation of pure water reported that the vapor side has higher temperature than the liquid side, while kinetic theories predictions have led to both possibilities. [7,20]

A basic theory for the mass transport at liquid-vapor interfaces was developed by Hertz [21] and Knudsen [22], and later extended by Schrage [23]. The Hertz-Knudsen theory assumes ballistic transport of vapor molecules between an evaporating and a condensing surface. Schrage considered the that the vapor leaving the surface has a drift velocity, leading to a modification of the Hertz-Knudsen expression. Subsequent efforts have been based on solving the Boltzmann transport equation (BTE), under the BGK approximation [9–11], or different approximations such as moment methods[24,25]. These efforts led to different approximate expressions relating the interfacial mass and heat fluxes to the liquid side pressure ($P_s$) and temperature ($T_l$), the vapor side pressure ($P_v$) and temperature ($T_v$), and the accommodation coefficient [26–28]. However, comparison of the kinetic theory predictions with experiments has not been successful [7,29]. Alternative theories had been developed, including nonequilibrium thermodynamics approaches [20,30] and statistical rate theory[31]. None of these approaches, however, were successful in explaining the experimental observations pioneered by Ward and co-workers [5,6,32]. It is also noteworthy that past efforts have focused almost exclusively on the vapor-phase transport, while very few studies had coupled the vapor-phase to the liquid-phase transport [33,34]. A recent work proposed maximizing interfacial entropy generation as a principle that can be used to determine the interfacial temperature jump, although the validity of such a principle can be questioned.[35]

Recently, the author of this work derived a set of interfacial conditions [36] by applying the diffusion-transmission type of approach, which has been used to treat molecular velocity slip in rarefied gas flow, photon and phonon interfacial discontinuities, and electron interfacial transport in the past [37]. In addition to the interfacial mass flux expression, which was identical to that derived by Schrage, a parallel expression was also given for the interfacial heat flux. These interface conditions can be coupled to transport in the vapor and the liquid phases to determine the interfacial temperatures on the liquid and the vapor sides, as well as transport in the liquid and the vapor phases.

In this work, the interfacial conditions derived before will be coupled with transport in both the liquid and the vapor sides of the interface to study the magnitude and the sign of temperature jump across the interface, as well as temperature profiles in both the vapor and the liquid phases. It will be shown that an intrinsic temperature jump always exists at the interface, caused by the difference in the enthalpy of the vapor molecules evaporating or condensing at the interface and in the bulk region. For evaporation, the vapor temperature near the interface is cooled below the saturation temperature on the liquid surface. For condensation, the reverse is true, i.e., the vapor temperature is heated above the saturation temperature on the liquid surface. For evaporation, due to the lower temperature of the vapor at the interface, however, reverse heat conduction from vapor away from the interface could happen, creating an inverted temperature profile in the vapor phase. When the supply of the heat from the liquid side is not sufficient, the reverse heat conduction from the vapor side can overwhelm the intrinsic cooling effect, leading to a higher temperature of the vapor phase than that of the liquid surface temperature. We will show that the model established here can explain past experiments. The temperature inversion had also been predicted before for the case of evaporation and condensation between two parallel plates [12,38]. The study carried out at a single interface provides physical picture for such paradoxical predictions. A more detailed study will be reported in another paper[39] for the case of evaporation and condensation between two parallel plates.

## II. MATHEMATICAL MODEL

### A. Interfacial conditions and governing equations in continuum

Ref.[36] started from the BGK approximation of the BTE and arrived at the following interface conditions for the mass and heat fluxes carried by the vapor across a liquid-vapor interface

$$m = \frac{2\alpha}{2-\alpha}\sqrt{\frac{R}{2\pi}}\left[\rho_s(T_s)\sqrt{T_s} - \rho_v\sqrt{T_v}\right] \quad (1)$$

$$q = \frac{2\alpha}{2-\alpha}R\sqrt{\frac{2R}{\pi}}\left[\rho_s(T_s)T_s^{\frac{3}{2}} - \rho_v T_v^{\frac{3}{2}}\right] \quad (2a)$$

$$= 2RT_s m + \frac{2\alpha}{2-\alpha}R\sqrt{\frac{2R}{\pi}}\rho_v\sqrt{T_v}(T_s - T_v) \quad (2b)$$

$$= 2RT_v m + \frac{2\alpha}{2-\alpha}R\sqrt{\frac{2R}{\pi}}\rho_s\sqrt{T_s}(T_s - T_v) \quad (2c)$$

where $\alpha$ is the accommodation coefficient, $\rho$ the density, R the ideal gas constant of the vapor, and T the temperature. The subscript "s" represents the properties of the saturated vapor phase on the liquid surface, and "v" the vapor phase properties at the outer edge of the Knudsen layer, i.e., $d_{w+}$ in Fig.1(a), which is of the order of a few mean free path lengths [40,41]. This thickness is neglected and hence "v" can be considered properties of the vapor phase immediately outside the liquid surface, as shown in Fig.1(a). It should be emphasized that the heat flux q in Eq.(2) does not equal to the wall heat flux supplied for evaporation, which also includes the energy needed to overcome the evaporation barrier, i.e., the latent heat. Similarly, in condensation, the wall heat flux includes the energy released as the vapor molecules descend from the vapor phase to the liquid phase. In the continuum vapor phase, i.e., outside the Knudsen layer, the same kinetic theory leads to local mass flux and the energy flux expression as

$$m \approx \rho u - R\rho T\tau \frac{dln(\rho T)}{dz} \tag{3}$$

$$q \approx \frac{5}{2}RT\rho u - \frac{5}{8}\tau(2RT)^2\frac{d\rho}{dz} - \frac{5}{4}\rho\tau(2RT)^2\frac{1}{T}\frac{dT}{dz} \tag{4}$$

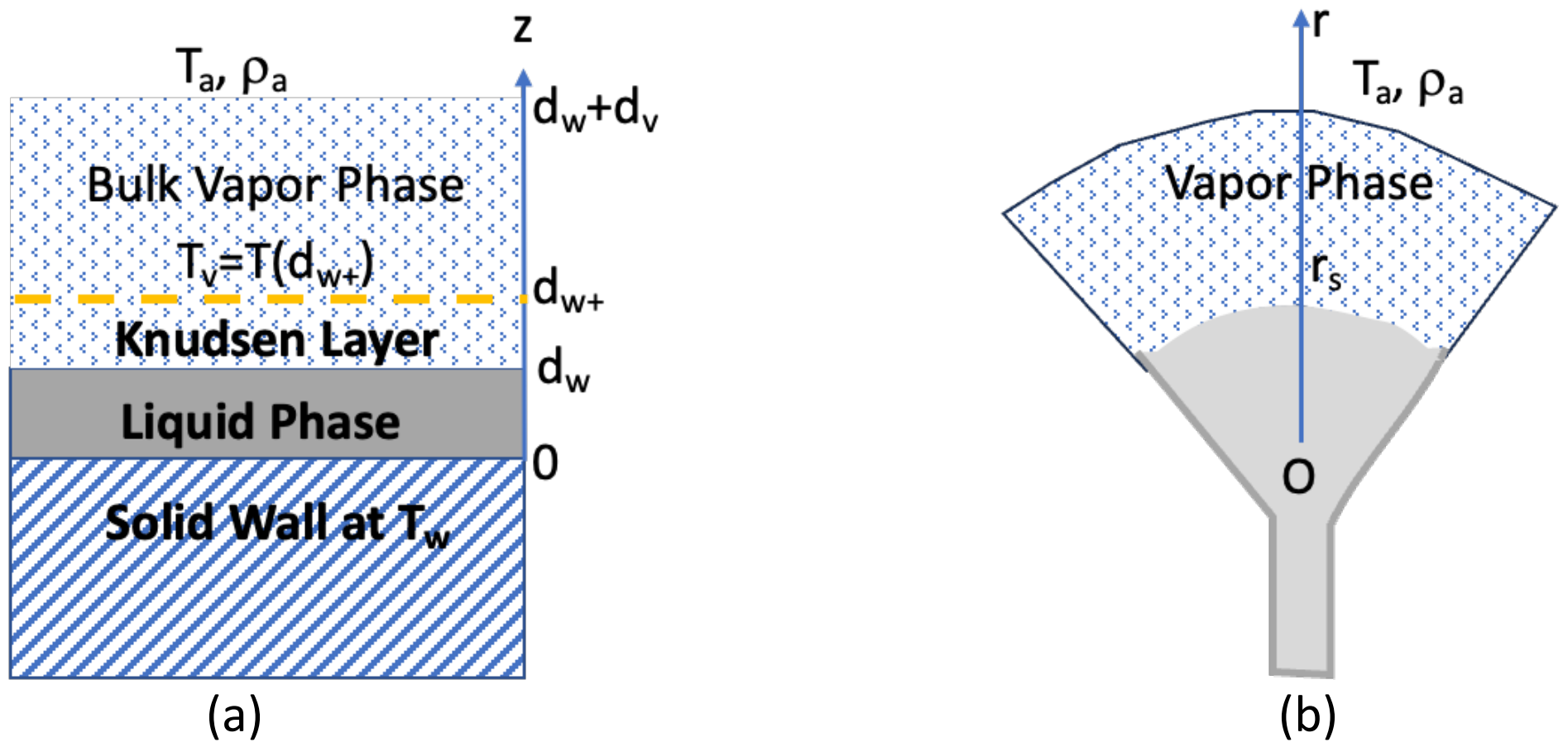

Figure 1 Coordinate used to study (a) planar geometry and (b) spherical geometry mimicking experimental configuration.

In Ref.[36], an additional relation for the interfacial vapor phase density $\rho_v$ was given. we realize now that this additional relationship is not needed. Because Eqs. (1) and (2) are flux expressions derived based on an approximate solution to the molecular distribution function, one should not further enforce the definition of thermodynamic properties based on the same approximations, which over-constrains the problem. We will see later that $\rho_v$ can be solved after specifying boundary conditions. Furthermore, in Ref. [36], we did not enforce the mass flux and velocity relationship

$$m = \rho u \tag{5}$$

arguing that the diffusion term in Eq. (3) contributes to mass flux. However, nearly all past approaches enforced this definition of velocity, with a few exceptions. For example, starting from the coupled mass and heat diffusion, Brenner argued that mass velocity is different from the momentum velocity[42,43]. He further introduced modification to the Navier-Stokes equation. Similar conclusions were reached using kinetic-theory based approach [44]. Fundamentally, it seems that a consistent approach requires modification of the momentum equations. In this work, however, we will assume Eq. (5) is valid, which when combined with Eq. (3) requires that

$$\rho T = Constant \tag{6}$$

If we assume the ideal gas law is valid, the above relation basically says that pressure in the vapor phase is a constant. This also means we do not need to solve the momentum equations, which was also the case in most of the past approaches based on solving the BTE. Using Eq.(6) to eliminate d$\rho$/dz in the energy flux, we get

$$q \approx \frac{5}{2}RTm - k\frac{dT}{dz} \approx c_pTm - k\frac{dT}{dz} \quad (7)$$

where the second equality replaces $\frac{5}{2}R$ with constant pressure specific heat $c_p$ so that the expression is applicable to polyatomic vapors. For monoatomic ideal gas, the thermal conductivity can be expressed as

$$k = \frac{5}{2}\tau(RT)^2\frac{\rho}{T} \quad (8)$$

We will again replace it with actual thermal conductivity of the vapor in numerical simulations for water. Eq. (7) can also be written into the familiar form of energy equation in the Navier-Stokes equation, assuming no heat is generated in the vapor stream, i.e., q = constant,

$$c_pm\frac{dT}{dz} = k\frac{d^2T}{dz^2} \quad (9)$$

where we have assumed that the thermal conductivity is temperature independent.

**B. Energy balance at a single interface**

Since the evaporation heat is supplied by the wall through the liquid layer [Fig.1(a)], heat carried by saturated vapor at the surface is

$$q = k_w\frac{T_w - T_s}{d_w} - m\Gamma(T_s) \quad (10)$$

where $\Gamma$ is the latent heat of evaporation at the saturation temperature $T_s$, $T_w$ the wall temperature, $d_w$ the liquid layer thickness and $k_w$ its thermal conductivity. The first term is heat conduction from the wall to the interface. The second term reflects the fact that in the kinetic theory formulation of interface conditions, the outgoing molecules from the interface already overcome the interfacial latent heat barrier [36]. Strictly speaking, the potential energy barrier to overcome during evaporation equals the internal energy difference between the liquid and the vapor phase, while the conventional latent heat is the enthalpy difference between the vapor and the liquid phase. Hence, $\Gamma$ should be replaced with $\Gamma' = \Gamma - \mathrm{p}_s v_s = \Gamma - \mathrm{R}T_s$. We used tabulated value of the latent heat for water in the results presented, and have checked that the difference between using $\Gamma'$ and $\Gamma$ for water is small, due to the large latent heat of water.

We point out that in most of the heat transfer analysis of evaporation and condensation, the wall heat flux is typically set to equal to the latent heat of the evaporated mass, i.e., q=0 in Eq. (10). In our treatment, the difference of the wall heat flux and the latent heat equals the heat transferred in the vapor phase, i.e., q given by Eq.(10) is the same q as given by Eqs. (2) and (4).

Although most of previous experiments did not use planar geometry, we will focus our discussion first using planar geometry, and specify the vapor boundary conditions at a distance $d_v$ away from the interface as [Fig.1(a)]

$$z = d_w + d_v \quad T = T_a, \ \rho = \rho_a \tag{11}$$

In experiments, typically the evaporating chamber pressure is given, i.e., $P = P_a$. Assuming ideal gas relation, we can relate $P_a$ to $\rho_a$ if $T_a$ is known. Solution to Eq. (10) with the above boundary conditions, and $T(d_{w+})=T_v$, is

$$T(z) = \frac{T_a[exp(c_p m(z-d_w)/k)-1] + T_v[exp(c_p m d_v/k) - exp(c_p m(z-d_w)/k)]}{exp(c_p m d_v/k)-1} \tag{12}$$

Using the above equation, the heat transported in the vapor phase can be calculated from Eq. (7) as,

$$q = c_p T_v m - \frac{c_p(T_a - T_v)m}{exp(c_p m d_v/k)-1} \tag{13}$$

where the first term represents convection and second term is due to conduction. The first term is always positive for evaporation ($m > 0$), and negative for condensation ($m < 0$). The second term can be either positive or negative. Equating the interfacial heat flux expression Eq. (2) with the heat flux expression Eq. (10), we have

$$2RT_v m + \Gamma m + \frac{2\alpha}{2-\alpha} R \sqrt{\frac{2R}{\pi}} \rho_s \sqrt{T_s}(T_s - T_v) = k_w \frac{T_w - T_s}{d_w} \tag{14}$$

We can further use Eq. (6) to write the interfacial mass flux expression as

$$m = \frac{2\alpha}{2-\alpha} \sqrt{\frac{R}{2\pi}} \rho_s \sqrt{T_s} \left(1 - \frac{T_a \rho_a}{\rho_s \sqrt{T_s T_v}}\right) \tag{15}$$

Applying the energy balance between the vapor phase Eq. (13) and the liquid phase Eq. (10), we get

$$k_w \frac{T_w - T_s}{d_w} = \Gamma m + c_p T_v m - \frac{c_p(T_a - T_v)m}{exp(c_p m d_v/k)-1} \tag{16}$$

Equations (14)-(16) can be solved simultaneously to obtain $T_s$, $T_v$, and $m$ for given $T_a$ and $\rho_a$ (or $T_a$ and pressure $P_a$). Due to the nonlinear nature of the $\rho_s$ and $T_s$ relationship, however, solving these combined equations is actually not straightforward and hence some details are given here. Substituting Eq. (15) into Eq. (14), re-arranging leads to

$$T_v + b\sqrt{T_v} + \frac{\Gamma}{2R} = 0 \tag{17}$$

where

$$b = -\frac{\rho_s\sqrt{T_s}}{T_a\rho_a}\left[\left(\frac{\Gamma}{2R}+T_s\right)-k_w\frac{T_w-T_s}{d_w\frac{4\alpha}{2-\alpha}R\sqrt{\frac{R}{2\pi}}\rho_s\sqrt{T_s}}\right] \tag{18}$$

Eq. (17) allows us to express $T_v$ explicitly in terms of $T_s$.

$$T_v = \left\{\frac{-b+\sqrt{b^2-4\frac{\Gamma}{2R}}}{2}\right\}^2 \text{ or } T_v = \left\{\frac{-b-\sqrt{b^2-4\frac{\Gamma}{2R}}}{2}\right\}^2 \tag{19}$$

We found that in most cases (when $\Gamma$ is large), the first expression leads to reasonable solutions, but in some other cases (when $\Gamma$ is small), we need to take the second expression. Substituting Eqs. (15) and (19) into Eq. (16) leads to an equation with the single unknown $T_s$, which can be readily solved using MATLAB. We can also nondimensionalize the above equations by introducing the following nondimensional variables

$$\Pi_1 = \frac{4\alpha\rho_a R\sqrt{RT_a}\,d_w}{\sqrt{2\pi}(2-\alpha)k_w},\ \ \Pi_2 = \frac{\Gamma}{RT_a},\ \ \Pi_3 = \frac{k_w/d_w}{k_v/d_v},$$

$$T_s^* = \frac{T_s}{T_a}, \rho_s^* = \frac{\rho_s}{\rho_a},\ T_w^* = \frac{T_w}{T_a},\ T_v^* = \frac{T_v}{T_a},\ m^* = \frac{\Gamma d_w}{k_w}\frac{m}{T_a},\ c_p^* = \frac{c_p}{R} \tag{20}$$

to recast Eqs. (15-19) into,

$$m^* = \frac{\Pi_1\Pi_2}{2}\rho_s^*\sqrt{T_s^*}\left(1-\frac{1}{\rho_s^*}\frac{1}{\sqrt{T_v^*T_s^*}}\right) \tag{21}$$

$$b^* = \frac{b}{\sqrt{T_a}} = -\rho_s^*\sqrt{T_s^*}\left[\left(\frac{\Pi_2}{2}+T_s^*\right)-\frac{T_w^*-T_s^*}{\Pi_1\rho_s^*\sqrt{T_s^*}}\right] \tag{22}$$

$$T_v^* = \left[0.5\left(-b^*+\sqrt{b^{*2}-2\Pi_2}\right)\right]^2 \tag{23}$$

$$T_w^*-T_s^* = m^*+c_p^*T_v^*\frac{m^*}{\Pi_2}-\frac{c_p^*(1-T_v^*)m^*/\Pi_2}{exp(c_p^*m^*\Pi_3/\Pi_2)-1} \tag{24}$$

Using Eqs. (21) - (23) as implicit expressions for $T_s^*$ allows one to solve for $T_s^*$ (and correspondingly $\rho_s$) with Eq. (24), and then $T_v^*$, and $m^*$, for given $T_w^*$ and nondimensional parameters $\Pi_1$, $\Pi_2$, and $\Pi_3$. Among these three nondimensional parameters, $\Pi_2$ is fixed for a given $T_a$, hence, only $\Pi_1$ and $\Pi_3$ should be considered as variables.

In solving the above equations, we need to specify the saturate temperature and density relation. We used the following correlation for saturated water vapor,[45]

$$\rho_s = 5.018 + 0.32321 \times t_s + 8.1847 \times 10^{-3} {t_s}^2 + 3.1243 \times 10^{-4} {t_s}^3 \quad (25)$$

where $t_s$=$T_s$-273 (°C). This fit works well for $t_s$ in [0, 40] °C. One can also use density-temperature relation based on the Clausius-Clapeyron equation, although accurate fitting also needs to include the temperature dependence of the latent heat. For density as the boundary z=$d_w$+$d_v$, we assume that the vapor density is related to its saturation density by a factor F, i.e.,

$$\rho_a = F\rho_s(t_a) \quad (26)$$

The value of F be determined by the pressure maintained during the experiment. F>1 means the vapor is supersaturated, while F<1 means it is undersaturated.

**C. Convection at liquid-vapor interface**.

For multidimensional problems or when the vapor side has additional forced motion, we can replace the heat conduction term in Eq. (7) with a convective boundary condition,

$$q \approx c_p T_v m + h(T_v - T_a) \quad (27)$$

where h is the convection heat transfer coefficient. Note that this h represents the single-phase heat transfer coefficient rather than the overall evaporative heat transfer coefficient. The latter can be calculated from the solved wall heat flux and the imposed temperature difference. Compared to results from previous section, we see that the only change we need to make is to replace $\Pi_3$ in Eq. (20) by

$$\Pi_3 = \frac{h}{k_w/d_w} \quad (28)$$

which is effectively the Biot number. Hence, we will not discuss convection case in this paper, which was treated more extensively in a previous paper.[36]

**D. Entropy Generation**

Since some of the temperature profiles presented later are counterintuitive, we will also calculate entropy generation to make sure that the transport process discussed does not violate the second law. We can apply the second law of thermodynamics to write the local entropy generation in the vapor-phase as [46]

$$\sigma = \frac{d}{dz}\left(\frac{q_c}{T}\right) + \frac{d(ms)}{dz} \quad (29)$$

where $q_c$ is the conduction heat flux, and s is the local entropy of the vapor. Since Eq. (6) implies that pressure is constant for the vapor phase, and the total energy flux in the vapor phase is a constant, we can write $q_c$ and ds as

$$ds = \frac{c_p}{T} dT \qquad \text{and} \qquad q_c = q - c_p T m \tag{30}$$

Using the above relations, we can express the local entropy generation in the vapor phase as

$$\sigma = -\frac{q_c}{T^2}\frac{dT}{dz} = \frac{k}{T^2}\left(\frac{dT}{dz}\right)^2 \tag{31}$$

which is clearly always positive. Hence, we do not need to worry about entropy generation in the vapor phase. Similarly, one can show that entropy generation for heat conduction in the liquid phase is also always positive.

To calculate the entropy generation across the interfacial region, we use the expression established by Bedeaux et al. [47], re-casted it into our notations

$$\sigma_i = -\frac{q(T_v - T_s)}{T_v T_s} - m\left[\frac{c_p(T_v - T_s)}{T_s} - \left(c_p ln\frac{T_v}{T_s} - R ln\frac{P_v}{P_s}\right)\right]$$
$$\approx -\frac{q(T_v - T_s)}{T_v T_s} - \frac{m(P_v - P_s)}{\rho_v\sqrt{T_v T_s}} \tag{32}$$

The first equation was derived from their original expression while the second one was the form Bedeaux et al. [47] used in developing the nonequilibrium thermodynamic transport theory for interfacial evaporation. We checked that the two expressions do not differ in numerical values.

## III. Results and Discussion

### A. Evaporation

**Inverted Temperature Profile.** In Fig.2a, we show temperature distributions in the liquid and the vapor phases for the same $T_w$ (=313 K) and $T_a$ (=298 K), but different F values in Eq. (26). The corresponding values of the vapor phase pressure, given by $P_a = R\rho_a T_a = P_v$ and saturation pressure, calculated by assuming ideal gas relation $P_s = R\rho_s T_s$, where R is the water's ideal gas constant, as well as the evaporation rates, are also given in the figure. The temperature distributions show rich behavior. When F=2, i.e., which means supersaturated vapor, the trend of temperature distribution is normal, decreasing from the wall to the liquid-vapor interface, and continuously decreasing from the vapor temperature at the interface to the vapor temperature at the boundary. However, at the interface, a temperature discontinuity occurs, dropping from the liquid side to the vapor side.

Even more interesting are the cases when F=1 and F=0.5. For such situations, the temperature profile is inverted. The interface temperature is lower than both the wall temperature and vapor phase temperature at the exit. Although this temperature seems to be strange at the first sight, it is consistent with our daily experience of evaporative cooling where water surface temperature[15,18,19] is colder than the ambient temperature. Again, a temperature jump happens across the interface, with the temperature of the liquid surface higher than the vapor side.

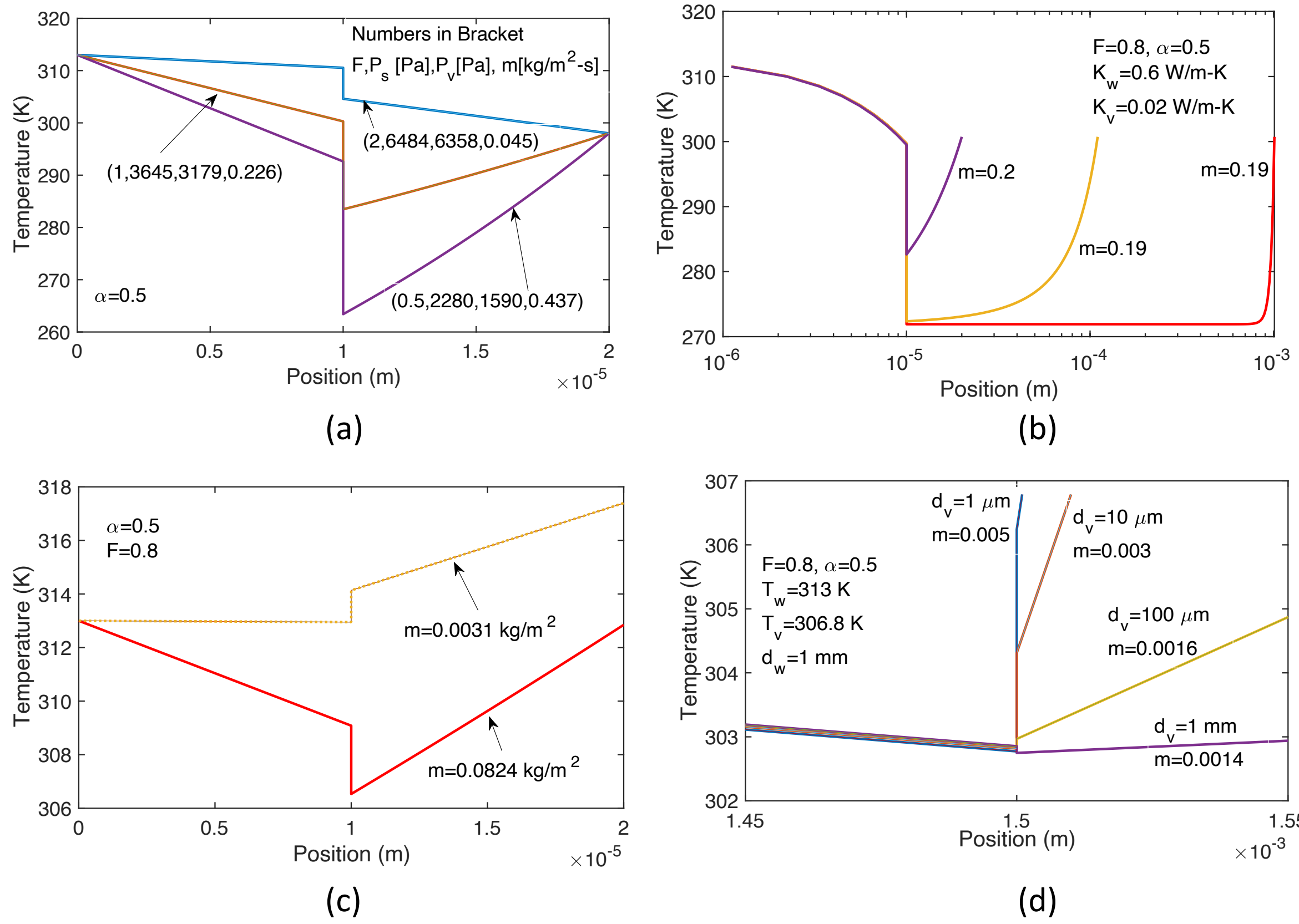

Figure 2 Temperature distributions in the liquid and vapor phases during evaporation, showing the impacts of different factors. (a) F values, with F=1 saturated, F>1 supersaturated, and F<1 undersaturated vapor at the boundary. (b) Vapor layer thickness, which determines reverse heat conduction. Larger vapor layer thickness leads to intrinsic interfacial temperature jump due to minimized reverse heat conduction and deeper cooling of the vapor phase. (c) Temperature at the vapor boundary. At higher vapor boundary temperature, large reverse heat conduction leads to a sign change of the interface temperature jump. And (d) A thick liquid layer with different vapor layer thickness, again showing interfacial temperature jump sign reversal. The following properties are used, close to liquid water and its vapor: $\Gamma$=2.45 MJ, $k_w$=0.6 W/m-K, $k_v$=0.02 W/m-K. Left of interface (where discontinuity happens) is water, and right of interface is vapor. Unit of the unmarked mass flow rate is [kg/$m^2$-s].

It should be kept in mind that in deriving the heat and mass flux interfacial conditions, i.e., Eqs. (1)&(2), the vapor side parameters are at the outer edge of the Knudsen layer, which is treated in the approximation as of zero thickness [Fig.1(a)]. Hence, one could think of interfacial temperature difference as the temperature drop across the Knudsen layer, although a small part of this temperature drop happens right at the liquid-vapor interface caused by the state-change, as indicated by molecular dynamics simulations [15,18,19].

**Intrinsic Interface Cooling Effect.** We can understand the lower vapor side temperature based on expression Eq. (2c) and bulk vapor-phase heat flux Eq. (13). First, we assume $T_a$=$T_v$, i.e., no heat conduction in the vapor phase. The convective heat in the vapor phase is the enthalpy carried, i.e., $\frac{5}{2}RT_v m$ for ideal gas, while in the interfacial heat flux expression Eq.(2c), the convective heat flux is $2RT_v\, m$. Setting this to Eq.(2), we have

$$\frac{2\alpha}{2-\alpha}R\sqrt{\frac{2R}{\pi}}\rho_s\sqrt{T_s}(T_s-T_v)=\frac{1}{2}RT_v m \tag{33}$$

where $\frac{1}{2}RT_v m$ comes from the difference of the enthalpy accompanying mass flow in the bulk region and the interfacial region. Equation (33) leads to $T_s \geq T_v$ for evaporation ($m \geq 0$) and $T_s \leq T_v$ for condensation ($m \leq 0$), i.e., cooling of the vapor during evaporation and heating during condensation. This temperature jump can be considered as the intrinsic. If $T_v \geq T_a$, the vapor phase heat flux will be larger than $\frac{5}{2}RT_v m$, and the interfacial temperature jump will be larger than the intrinsic value. On the other hand, if $T_v \leq T_a$, the vapor phase heat flux will be smaller than the convective value due to reverse heat conduction. When the reverse heat conduction contribution is more than $\frac{1}{2}RT_v m$, the interfacial temperature drop reverses its sign, i.e., the vapor side will be hotter than liquid side.

Figures 2(b)-2(d) illustrate these points. In Fig.2(b), because the liquid film is thin (10 μm), the latent heat for evaporation is mainly supplied by the liquid side, leading to high mass flux. The interfacial temperature jump is largest when $d_v$ is large because in this case, the reverse heat conduction is small. As the vapor layer becomes thinner, the temperature jump across the interfaces diminishes due to the reverse heat conduction in the vapor phase. In Fig.2(c), we show the reversal of the interfacial temperature difference when the vapor side is also thin and the vapor temperature at the boundary is higher than the wall temperature. This situation is further amplified in Fig.2(d), when the liquid film becomes thick, and the reverse heat conduction from the vapor side supplies a significant amount of heat needed to overcome the latent heat of evaporation. In this case, the vapor side temperature is higher than that of the liquid side.[47] Past experiments measuring liquid-vapor interfacial temperature jumps have shown large variations [2,8]. The above results shed lights on why there are such variations. In Sec. V, we will present comparisons of the model results with some of the past experiments.

**Heat Flux Saturation.** Figure 3(a) shows nondimensional interfacial temperature as a function of $\Pi_1$ at different $\Pi_3$ values. A small $\Pi_1$ value means the liquid film is thin. A small $\Pi_3$ value means the vapor side is thin. We can see that the liquid interface temperature does not change much with $\Pi_3$, because heat conduction in the liquid pins the liquid side temperature close to the wall value. However, the vapor phase temperature changes significantly with both $\Pi_1$ and $\Pi_3$ values.

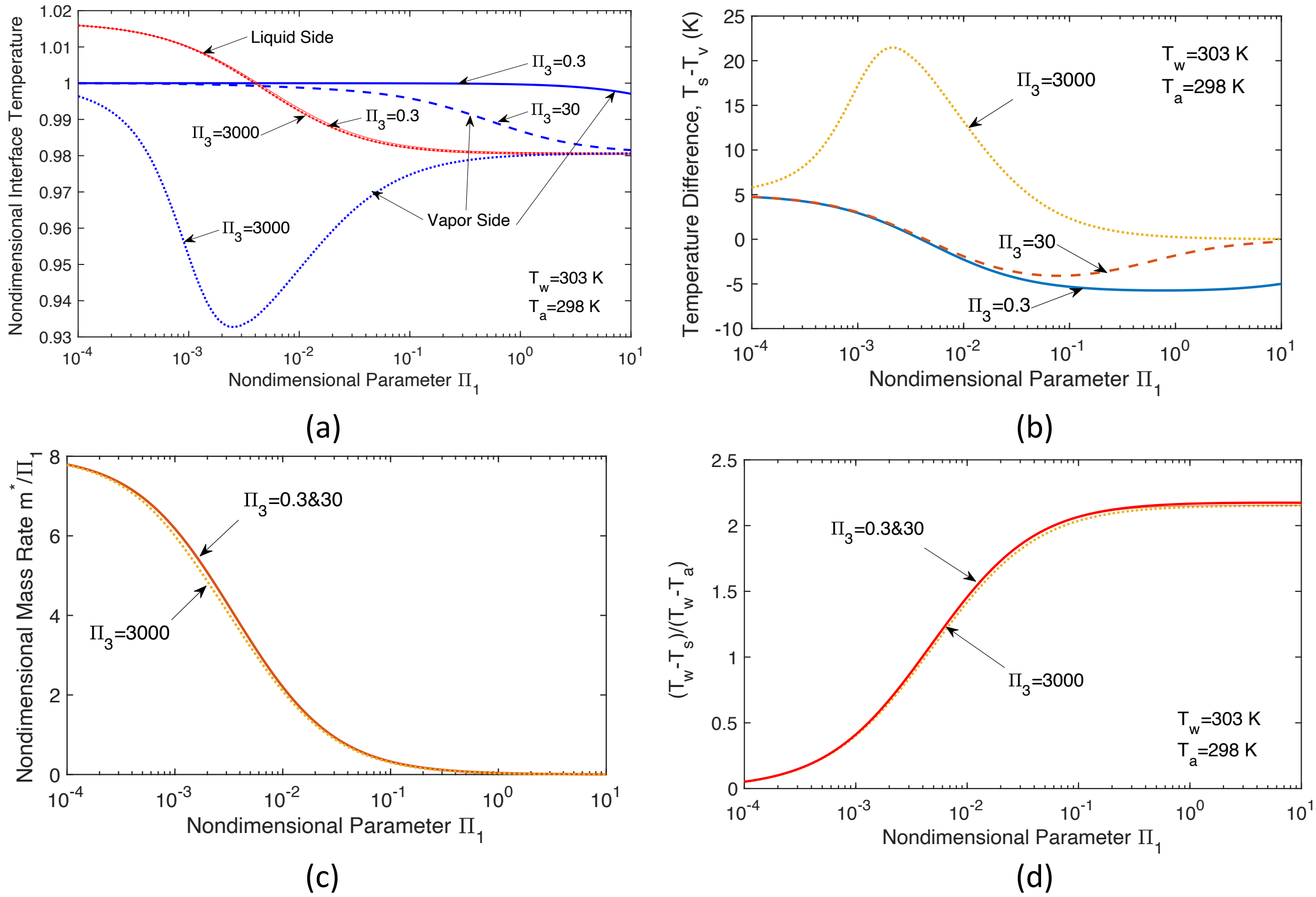

Figure 3 Dependence on nondimensional thickness $\Pi_1$ at different $\Pi_3$ values. (a) Nondimensional interfacial temperature on the liquid and the vapor sides, (b) temperature difference at the interface, (c) nondimensional evaporative rate, and (d) nondimensional temperature drop across the liquid film.

When $\Pi_3$ is large such that most heat supplied is from the liquid side, the vapor side temperature could be much lower that the liquid side due to the intrinsic interfacial cooling effect. As $\Pi_3$ becomes smaller, reverse heat conduction from the vapor side to the interface becomes significant, which can eventually lead to a reversal of the sign of the interfacial temperature difference, as shown in Fig.3(b). Figure 3(c) shows the $m^*/\Pi_1$ as a function of $\Pi_1$. We choose to present $m^*/\Pi_1$ since this combination eliminates the liquid film thickness in nondimensional $m^*$. We observe that $m^*/\Pi_1$, i.e., the evaporation rate m, does not change much with $\Pi_3$, due to the much smaller thermal conductivity of the vapor phase. As the liquid film gets thinner, i.e., $\Pi_1$ becomes smaller, the evaporation rate approaches an asymptotic value. This is contrary to most of the past retreatments on evaporation which neglect the interfacial temperature drop, leading to an evaporation rate, and the corresponding heat transfer rate, inversely proportional to the liquid film thickness. Figure 3(d) shows the temperature drop across the liquid film vs. the overall temperature difference applied for evaporation. As $\Pi_1$ becomes smaller, i.e., the liquid film becomes thin, the temperature drop across the liquid film becomes smaller. Most of the applied temperature drop between the wall and the vapor phase happens between the liquid and the vapor interface. This explains why the evaporate rate saturates at small $\Pi_1$ values. Also, we notice that the nondimensional temperature drop across the liquid film saturates to twice of the applied temperature difference, reflection the fact of evaporative cooling at the interface.

**Interfacial Entropy Generation.** We have checked that the peculiar behavior of the vapor phase temperature and interfacial temperature discontinuities does not violate the 2$^{nd}$ law of thermodynamics by calculating the entropy generation across the interface according to Eq. (32). Fig. 4(a) shows the interfacial entropy generation as a function of the temperature difference between $T_w$ and $T_a$. The corresponding mass flux and interfacial heat flux are plotted in Fig.4(b), but as a function of the calculated interfacial temperature jump $T_s$-$T_v$. We can see that the interfacial entropy generation for all cases is positive. Evaporation is a mass transfer process, which is driven by the chemical potential difference. Figure 4(b) shows that the interfacial temperature differences ($T_s$-$T_v$) changes sign even when m is still positive. There is also a region where m is positive, but q is negative, i.e., the reverse heat conduction is larger than the enthalpy carried forward by the evaporating heat flux such that q is negative. In Fig.4(c), we show the density distributions corresponds the two cases in Fig.2(c). The saturated vapor density at the liquid surface is exaggerated by plotting the density into the liquid. Across the interface, the density of the saturated vapor on the liquid side is larger than that of the vapor side, despite that the corresponding temperature difference at the interface reverses sign as shown in Fig.2(c). In Fig.4(d), we show both the temperature and density distributions for a set of conditions given in the figure. Despite that the vapor phase density gradient is towards the interface, the density on the liquid side is still larger than on the vapor side at the interface. Such results are consistent with the fact that mass transfer is mainly driven by the chemical potential difference (although it could be coupled to the temperature difference),[46] which is directly related to vapor densities at the two sides of the interface. One can appreciate the cooling effect as resulting from the sudden expansion of the vapor from the saturated state on the liquid surface to a lower density at the outer edge of the Knudsen layer, similar to the Joule-Thomson expansion.[48] More discussion on the similarities and differences of the interfacial cooling to Joule-Thomson effect will be made later.

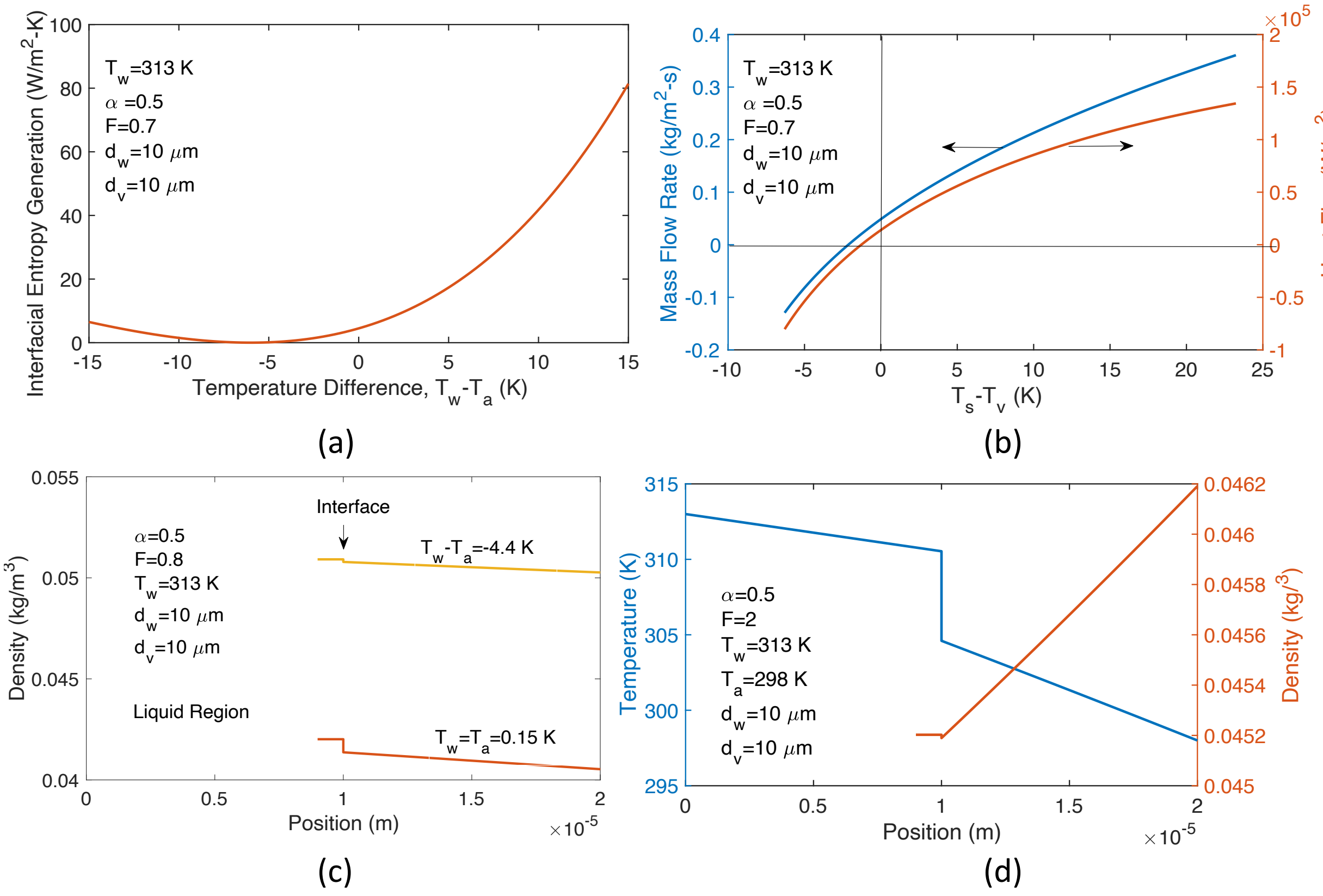

Figure 4(a) Interfacial entropy generation vs. applied temperature difference, showing it is always positive for both condensation and evaporation, (b) mass flux and heat flux vs. calculated interfacial temperature jump, (c) interfacial density discontinuity for cases corresponding to Fig.2(c), and (d) temperature and density distributions. The region of saturated vapor density at the liquid surface is exaggerated by plotting the density into the liquid.

## B. Condensation

**Temperature Inversion and Interfacial Heating Effect.** The same set of equations can be used to study condensation, as indicated by the positive entropy generation in Fig.4(a). In Figure 5(a), we show temperature distributions when $T_w$=313 K and $T_a$=298 K for evaporation, and $T_a$=313 K

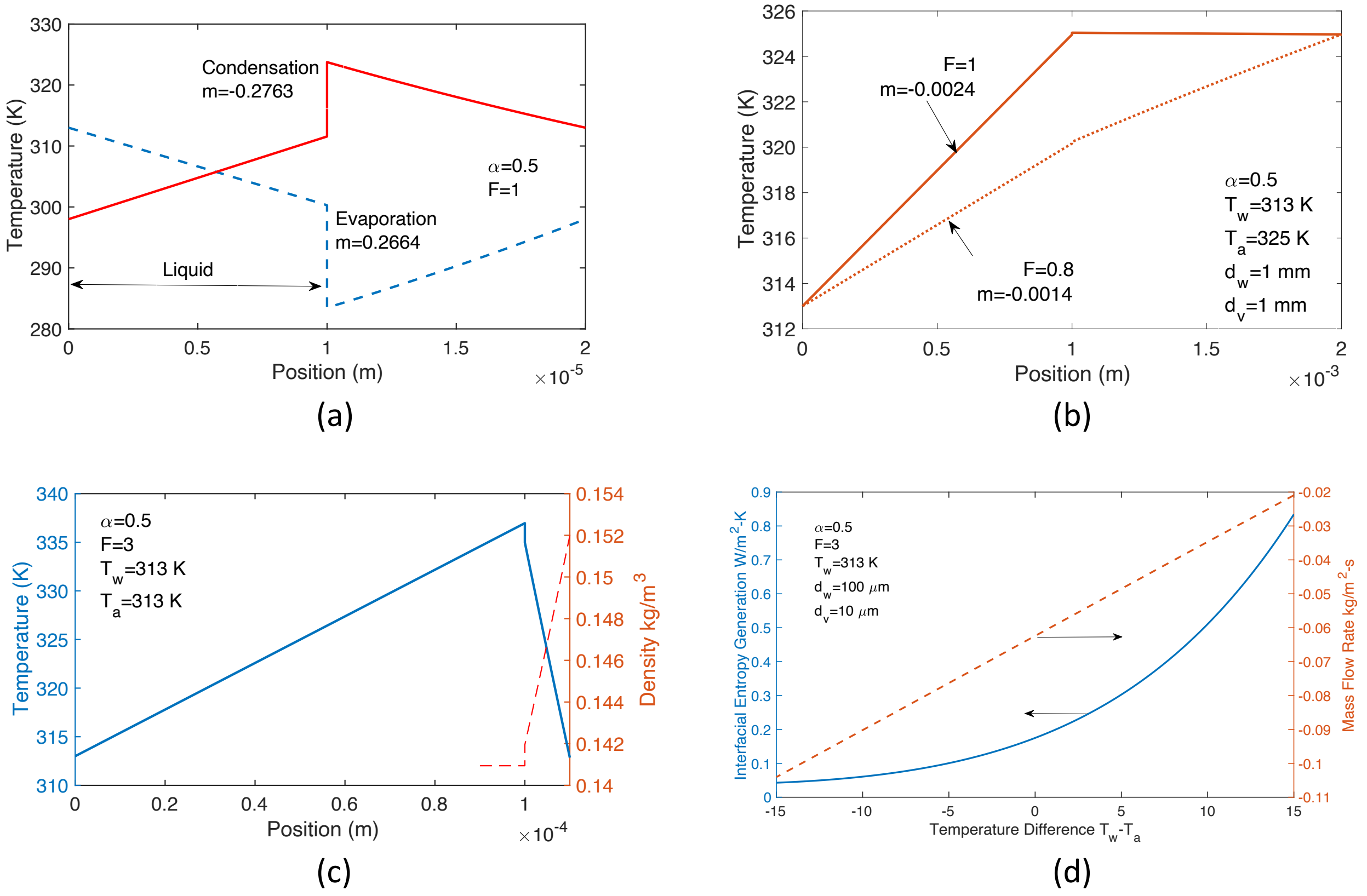

Figure 5 (a) Comparison of temperature profiles in the vapor and liquid phases during evaporation and condensation, for same applied temperature difference, (b) temperature profile when both liquid and vapor phases are thick, (c) inverted temperature profile with liquid side ks at higher temperature than the vapor side, and (d) entropy generation and the corresponding mass flow rate with a negative sign meaning condensation.

and $T_w$=298 K for condensation. Similar to evaporation, condensation can also have an interfacial temperature drop and can create an inverted temperature profile with the interface having apparently higher temperature than both the wall and the vapor temperature at the boundary, as shown in Fig. 5(a). In this case, the enthalpy carried by the vapor coming towards the interface is 5Rm/2, while that of molecules condense is 2Rm. The imbalance leads to vapor heating. One can also think that the incoming vapor is at a higher density than that of saturated vapor on the liquid surface, i.e., the vapor cools down upon condensation. This vapor-phase heating phenomenon is opposite to the intrinsic cooling effect in evaporation, and consistent with the requirement of Eq. (33) since m is now negative. Despite the similarities, however, evaporation and condensation temperature profiles are not exactly symmetric. Evaporation has a larger interfacial discontinuity (16.8 K) than condensation (12.2 K) as shown in Fig.5(a). In Fig.5(b), we show a more normal condensation temperature profile when both the liquid and the vapor phase are thick, for F=0.8. In this case, the liquid layer is thick, and hence the condensation rate is low.

In condensation, heat is released during condensation in the liquid side. Reverse heat conduction to the boundary at $z=d_w+d_v$ in the vapor phase can effectively help dissipate heat generated at the interface, which can lead to a reversal of sign of the interfacial temperature jump, as shown in Fig.5(c). In this case, the vapor layer is much thinner, pulling away more heat flux directly from the liquid side. In Fig. 5(d), we again show that the entropy generation at the interface is always positive, consistent with the second law. The case presented in Fig.5(c) corresponds to $T_w$-$T_a$=0 in Fig.5(d), which of course has a positive entropy generation. Similar to evaporation, condensation is mainly driven by the density difference, and the vapor side density is higher than that of saturated vapor on the liquid surface in all cases studied.

**Heat Flux Saturation.** We plot in Fig. 6(a) the nondimensional surface temperature on the liquid and the vapor sides as a function of $\Pi_1$ at different $\Pi_3$ values. The liquid surface temperature is quite independent of $\Pi_3$, and it approaches wall temperature when $\Pi_1$ is small, again due to efficient heat conduction in the liquid. Similar to evapoation, the vapor side temperature has much large variations. Figure 6(b) shows the interfacial temperature difference from the vapor-side to the liquid side. Figure 6(c) gives the nondimensional condensation rate, which agains saturates as $\Pi_1$ becomes small, because when the liquid film is thin, most of the temperature drop happens across the vapor-liquid interface, as shown in Fig. 6(d). Unlike the evaporation case, however, the nondimensonal liquid surface temperature approaches one for the conditions given. We do point out, however, this is not the general case as suggested in Fig.5(c).

**C. Metastable complete temperature inversion cases**

In Figs. 7, we show a condensation and an evaporation cases that are completely against the intuition, but are allowable solutions. Fig.7(a) is the case of condensation, but the temperature continuously drops from the wall to the vapor phase. However, the density is higher in the vapor phase and mass transfer occurs from the vapor to the liquid phase, i.e., condensation happens, as shown in Fig. 7(b) where the mass transfer rate for the condition in Fig.7(a) ($T_w$-$T_a$=5 K) is negative. The same figure also shows that the entropy generation at the interface is also positive. In this case, the vapor is undercooled below the saturation temperature at the interface. Since the vapor phase is very thin, condensation heat released at the interface conducts to the vapor

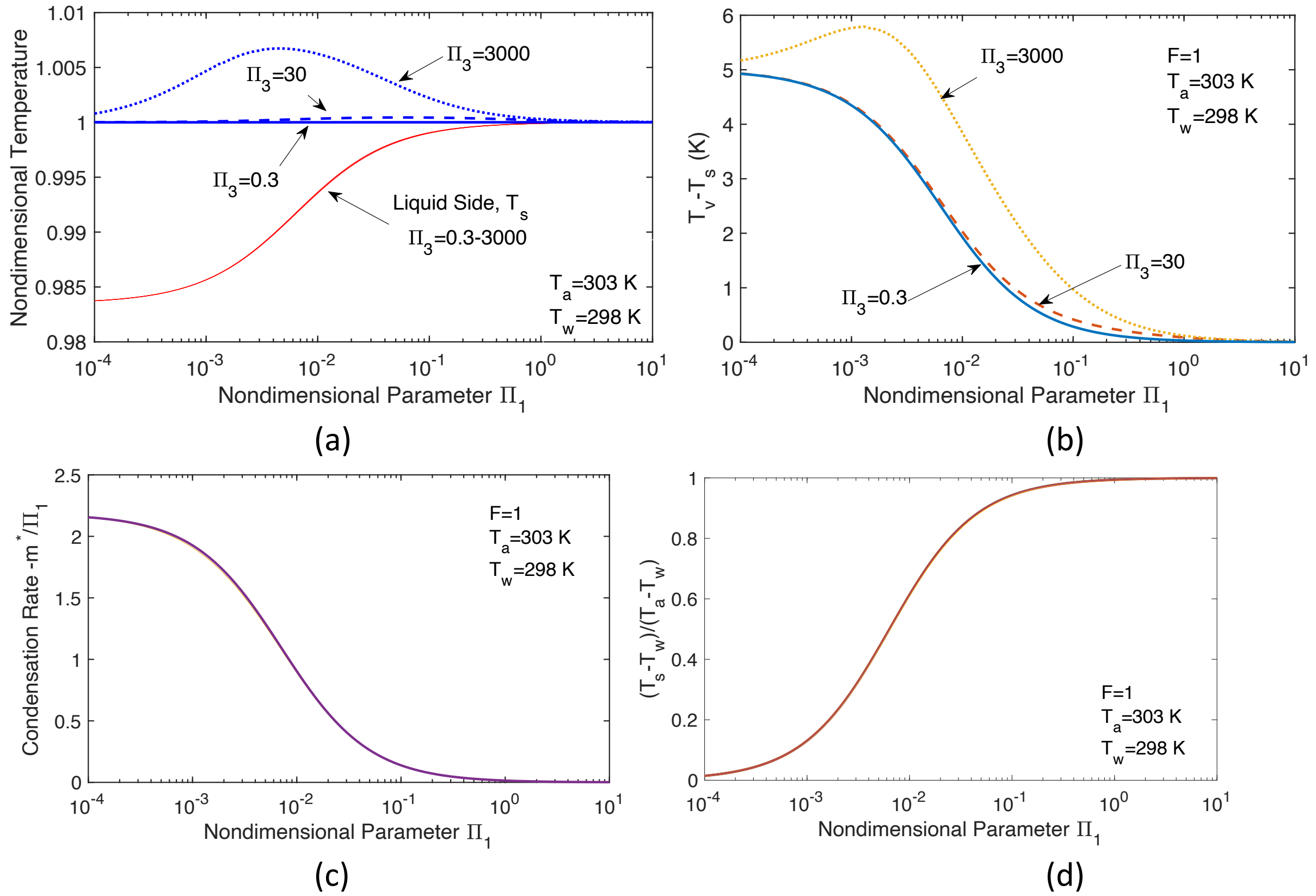

**Figure 6** (a) Nondimensional temperatures at the liquid and the vapor side, (b) interfacial temperature drop from the vapor to the liquid side, (c) nondimensional condensation rate, and (d) nondimensional temperature drop across the liquid film normalized to total temperature difference applied.

boundary. In fact, since the condensation rate is low, even the wall conducts heat to the vapor boundary. Such a situation is nearly impossible to create in reality because the undercooled vapor is likely to condense at the vapor boundary rather than at the interface. Hence, we call this as meta-stable. Figure 7(c) shows an opposite evaporation case, where the vapor phase is at higher temperature than the saturation temperature, but lower density than the saturation density. The mass transfer at the interface is from the liquid to the vapor, as shown in Fig.7(d) ($T_w$-$T_a$=-5 K), and the entropy generated at the interface is again positive. In this case, heat supplied from the vapor is more than enough to evaporate water, the excess conducts to the wall. This situation is again difficult to create as the low-density vapor will likely to condense at the boundary rather than at the interface.

## VI. COMPARISON WITH EXPERIMENTS

Reported experiments on the temperature jump at interfaces vary significantly among different groups [14,32,34,49–52]. With the exception of [52], all other experiments reported that the vapor phase has a higher temperature than the liquid phase. Measuring temperature distributions across liquid-vapor interface is prone to error, which might have contributed to the over two-orders-of-magnitude variations in the reported interfacial temperature discontinuities [8]. However, our model also suggests that the interfacial temperature discontinuity depends sensitively on the accommodation coefficient, the liquid and vapor phase boundary conditions and thickness, and experimental geometry.

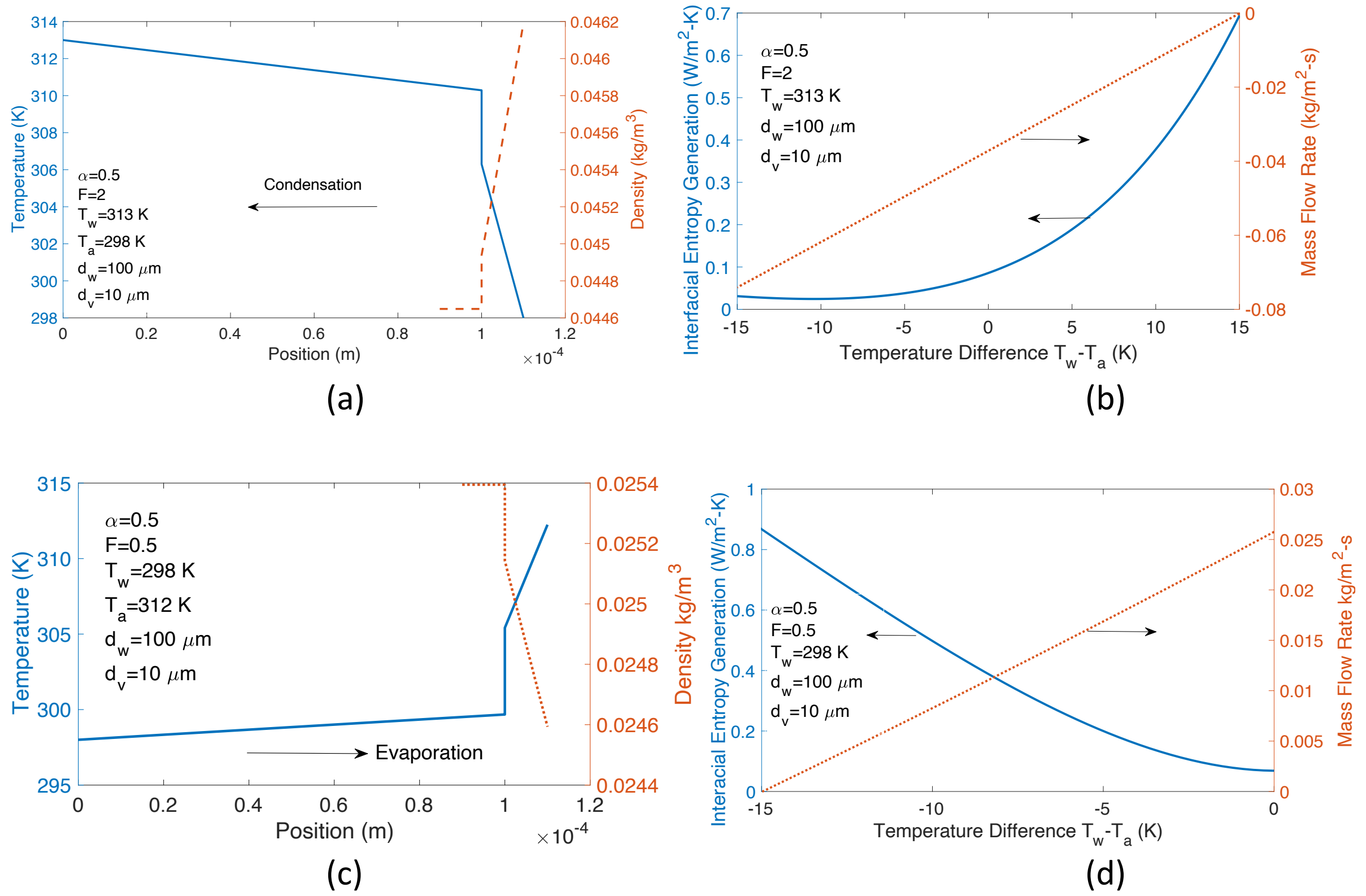

Figure 7 Complete temperature inversion. (a) Condensation happens even when the temperature drops from the wall to the vapor phase, (b) entropy generation at interface is still positive and mass flow rate is negative (condensation). (c) Evaporation happens even when the temperature drops from the vapor phase to the wall, (d) entropy generation at the interface is positive and mass flow rate is positive (evaporation).

Ward and co-workers carried out pioneering experiments measuring the temperature discontinuities during evaporation and condensation. Past modeling to explain their experimental data had not been successful [33]. We choose a set of experiments by Ward and Stanga [50] since they include both evaporation and condensation. Their experiments employed a conical funnel to hold water. The vapor phase temperature is not artificially controlled. Hence, the chamber walls serve as the boundaries setting the vapor phase temperature. To model their experiment, we approximate the transport of the evaporated water vapor as spherically symmetric flow [Fig.1(b)], while keep in mind that diameter of their vapor chamber is smaller than the height on top of the water cup, and hence the vapor phase transport is not exactly spherical. The vapor phase equation can be written as

$$Q \approx c_p T M - k 4\pi r^2 \frac{dT}{dr} \tag{34}$$

where Q is the total heat transfer and $M$ is the total mass flow rate. The solution of the above equation is

$$T(r) - \frac{Q}{c_p M} = \left(T_v - \frac{2Q}{5RM}\right) exp\left[-\frac{c_p M}{4\pi k}\left(\frac{1}{r} - \frac{1}{r_s}\right)\right] \tag{35}$$

Since the above temperature asymptotically approaches the ambient temperature, we will neglect the finite distance between the water cup and the chamber wall and apply the boundary condition that when $r \to \infty$, $T \to T_a$, which leads to

$$Q = c_p M \, \frac{T_v exp\left(\frac{c_p M}{2\pi k r_s}\right) - T_a}{exp\left(\frac{c_p M}{2\pi k r_s}\right) - 1} \tag{36}$$

Or, in terms of heat flux,

$$q = c_p m \frac{T_v exp\left(\frac{c_p m r_s \phi}{k}\right) - T_a}{exp\left(\frac{m r_s \phi}{k}\right) - 1} \tag{37}$$

where $\phi$ is the fraction of the spherical area of the evaporating surface. We still have the relation between density and temperature, $\rho T = Constant$.

For the liquid side, we also tried to model the funnel with a spherical geometry. We estimate that for the same surface heat flux, the equivalent wall thickness for such a spherical geometry to that of a parallel plate is $d_w \sim 40\ mm$. However, heat conduction along the funnel wall, which seems to be made of stainless steel, makes the spherical geometry assumption questionable. Instead, we treat heat conduction in liquid using the planar geometry model with $d_w$ adjustable. Hence, we only need to change the energy balance Eq. (16) as

$$k_l \frac{T_w - T_s}{d_w} = \Gamma m + c_p m \, \frac{T_v exp\left(\frac{c_p m r_s \phi}{k}\right) - T_a}{exp\left(\frac{c_p m r_s \phi}{k}\right) - 1} \tag{38}$$

We also set $r_s$ equaling the radius of the funnel mouth, $r_s$=3.5 mm. In Fig.8, we show comparison of modeled result with the experimental data for both evaporation and condensation. The parameters were adjusted for evaporation. Results for condensation is based on the same parameters, except applying the experimentally measured temperature at z=-5 mm of the liquid side as $T_w$ and using the pressure of the vapor phase given in their experiment.

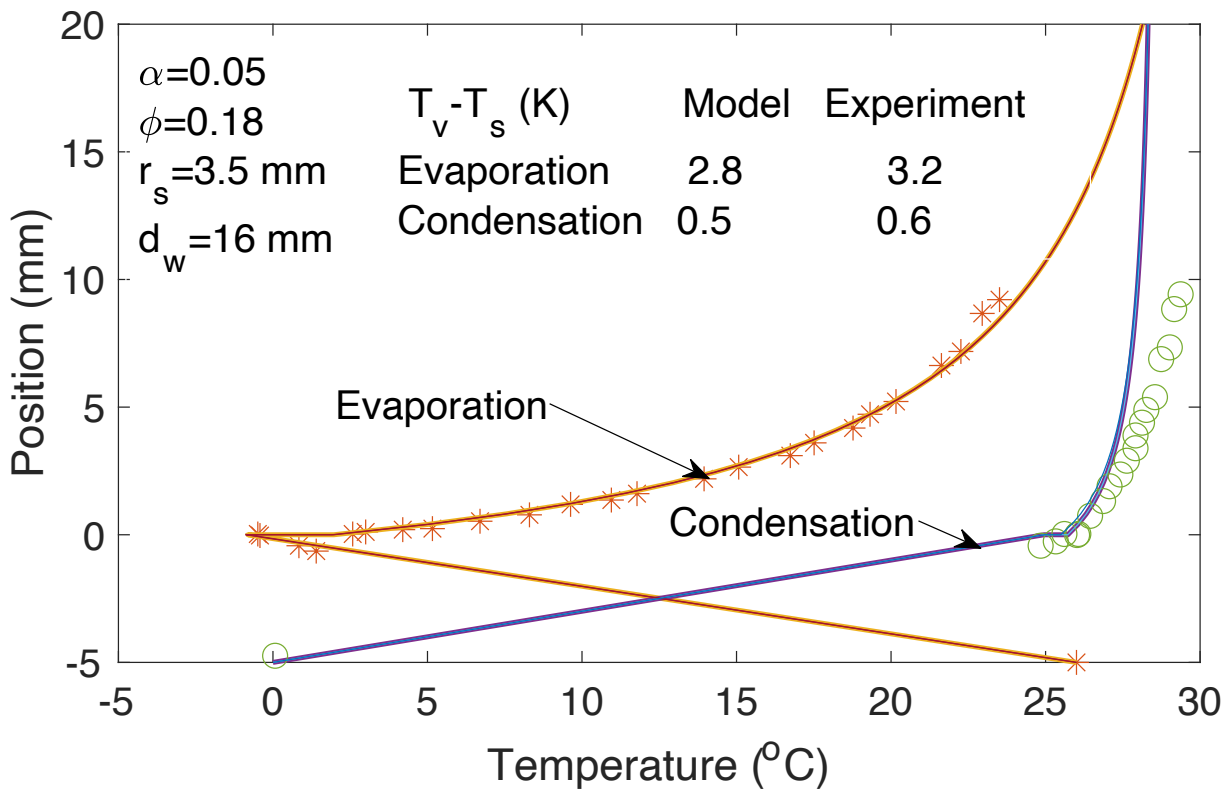

Figure 8 Comparison between modeling and experimental data for evaporation and condensation.

We see that the current model can well capture the asymmetry in the temperature profiles between the evaporation and the condensation. The model can also well explain the experimental data in the interfacial temperature jump. For the condensation case, the modelled condensation rate is 6.9 μg/s while experimental data is 12.7 μg/s. The difference in the rate of evaporation is larger, the modelled rate is 9.3 μg/s while experimental data is 41.9 μg/s. Considering the uncertainties in the experimental conditions, we view that the model and experiments come to the same order of magnitude a good validation of the model. Our model reveals that the reason that the evaporation experiment reported a higher vapor phase temperature is because the liquid layer is too thick. On the other hand, in the condensation experiment, the higher vapor phase temperature is consistent with prediction by the interfacial intrinsic cooling effect, although the interfacial temperature drop is much smaller than observed in the evaporation experiment.

## V. DISCUSSION

It is interesting to compare the intrinsic interfacial cooling effect during evaporation with several other familiar effects.

**Comparison with Joule-Thomoson Effect in Fluids.** In Joule-Thomson effect,[48] a gas or liquid goes through a sudden pressure drop in an adiabatic process, so that the enthalpy remains the same. Due to the pressure work PV (where V is the specific volume) value changes, the internal energy of the fluid changes, leading to a change in the fluid temperature. The Joule-Thomson coefficient, defined as $\mu_{JT} = (\partial T/\partial P)_H$ for many fluids is positive near ambient conditions, which leads a cooling effect. Accompanying the Joule-Thomson expansion is often a decrease of the vapor density, as in the Joule-Thomson throttling valve used in refrigerators. The density reduction at the interface for both evaporation and condensation bears similarity to the sudden expansion in a throttling valve and hence to the Joule-Thomson effect. However, the intrinsic cooling effect at the interface is not created in an isentropic process. The equivalent specific enthalpy at the interface associated with the mass flow is $2RT_v$ (for monotonic ideal gas) according to Eq.(2c), while in the bulk phase is $2.5RT_v$ according to Eq.(13). This change in enthalpy leads to a cooling effect on the vapor at the interface. The second term in Eq.(2c) supplies heat from interface to vapor to ensure energy balance.

**Comparison with Thomson Effect in Solid.** Thomson effect is due to the temperature dependence of the thermal energy carried by electrons when they conduct along a solid of nonuniform temperature.[53] If we think the thermal energy as the enthalpy of electrons, the above-mentioned enthalpy difference across the interface leading to the intrinsic cooling and heating effects during evaporation and condensation, respectively, bears similarity to the Thomson effect. However, the Thomson effect is continuous, typically slow changing, while the interfacial enthalpy change happens rapidly over the Knudsen layer, leading to large cooling or heating at the interface for evaporation or condensation, respectively.

**Comparison with Thermionic Cooling**. Thermionic emission of electrons from an interface also leads to cooling.[54] The cooling power per electron is $(\Delta + 2k_BT)$, where $\Delta$ is the barrier height and $k_B$ the Boltzmann constant. For vapor, the outgoing cooling power per kilogram of vapor is $(\Gamma' + 2RT)$, identical to that of electrons. This is consistent with similarities of evaporation to thermionic emission. However, in thermionic cooling, either vacuum or solid-state barriers are considered,[54,55] while in evaporative cooling modeled here, the Knudsen layer is followed by a continuum transport region including convection and conduction.

**Comparison with Coupled Thermionic and Thermoelectric Transport.** The evaporative cooling effect studies here is more similar to a previous treatment by the author[56] on electron transport across an interfacial potential step followed by thermoelectric transport in the bulk region. In that case, the thermoelectric current driven by a temperature gradient cannot be supplied sufficiently by an electron emitting region, leading to an electron temperature drop at the interface and the corresponding voltage drop. In the present case, the enthalpy carried away by convection in the bulk region cannot supplied by the enthalpy flux emitted at the interface during evaporation, leading to the interfacial temperature drop.

**More on the Microscopic Picture.** Why the enthalpy of a monoatomic gas changes from $2RT_v$ for emission at the surface to $2.5RT_v$ for convection in the bulk phase? Both results are rigorously derived from the kinetic theory.[36] This change can be understood as arising from the difference in the average velocity of molecules. On the vapor surface, the emitted vapor molecules obey the Maxwellian velocity distribution, with a zero average velocity, while in the outer edge of the Knudsen layer, the average molecular velocity is nonzero, and a displaced Maxwellian velocity distribution was assumed to be the local equilibrium distribution. One can thus connect this enthalpy change to the velocity change. More details insights could be obtained by examining the full solution of the Boltzmann equation through the Knudsen layer.

**Limitations of the Current Model.** The above discussions immediately raise some questions on if the current model is valid when $d_v$ is smaller than the Knudsen layer, i.e., at large Knudsen numbers. We believe that the trend will be similar although the values may change, for the following reasons. First, most of the past solutions of the BTE evoked the BGK approximation with the displaced velocity distribution. It was found that the difference between exact solution of the BTE and the BGK approximation is usually small.[57] Second, similar problems had been extensively studied for photon and phonon thermal transport between two parallel plates, based on solving the BTE or using discontinuous boundary conditions coupled to the diffusion approximation.[41] It was found solutions based on these two approaches differ very little spanning from small to large Knudsen numbers.

Another question one can raise is if the $2RT_vm$ and $2.5RT_vm$ difference of $0.5RT_vm$ between the enthalpy of molecules emitted at the surface and the bulk region is always valid, since these results are derived for monoatomic gas. Of course, $c_p$ of polyatomic gas will be different for polyatomic gases. Past treatments of the polyatomic gas have added energy associated with other internal degrees of freedom to the Boltzmann-Maxwellian distribution, while maintaining the kinetic energy term the same.[58] Such treatment would lead to the conclusion that the difference of enthalpy remains 0.5R. In our calculation, the second equation of Eq.(7) used $c_p$ of water, while the interface emission enthalpy remained $2RT_vm$. Hence, the exact numbers presented may change if such detailed are considered, but the trend should remain similar. Self-consistent treatment of polyatomic gas is needed to fully examine this problem, including the boundary conditions.

Finally, we note that this treatment has not included noncondensable gas. For evaporation in air and other relevant applications, the effect of noncondensable gas need to be included. Extension of the work to noncondensable gas should be straight forward, following established strategies.[58–60]

## VI. CONCLUSION

In the above, we have shown that when evaporation or condensation happens, an intrinsic temperature difference develops across the interface. In evaporation, the vapor-phase temperature is lower than the saturated vapor temperature at the liquid surface, i.e., the vapor is cooled. In condensation, the vapor-phase temperature is higher than the saturated vapor temperature at the liquid surface, i.e., the vapor is heated. This intrinsic interfacial phenomenon is due to the difference in the enthalpy carried by the vapor in the bulk region and molecules emitted or absorbed at the interface. Heat transfer created by the interfacial temperature difference at the interface compensates the enthalpy difference, to ensure the continuity of the energy flux across the interface and the bulk region.

The success in modeling is built on using the recently derived interfacial heat flux condition for evaporation and condensation with the well-established Hertz-Knudsen-Schrage theory for interfacial mass flux to bridge the transport in the liquid and the vapor continuum. The solutions show rich behavior of the temperature profiles. The temperature gradient in the vapor phase can be along or against the mass flow direction. In the former case, inverted temperature is created in the vapor phase. We also predict the possibility that the temperature gradients in both the vapor and the liquid phases are opposite to conventional wisdom. Entropy generation calculations show that the counter-intuitive temperature profiles do not violate the second law, since evaporation and condensation are fundamentally mass transport process driven by chemical potential differences.

Although the enthalpy mismatch leads to higher liquid side temperature for evaporation and lower for condensation intrinsically, the reverse heat conduction from the vapor phase can diminish the intrinsic interfacial cooling and heating effect, even reversing the sign of the interfacial temperature jump. Our modeling results could reasonably explain existing experimental data reported in the past on interfacial temperature drops for evaporation and condensation, which had defied modeling efforts in the past. Most of past experiments for evaporation reported opposite trend of the interfacial temperature jump. We explain this difference through modeling as arising from the reverse heat conduction in the vapor phase, especially these experiments used a thick liquid layer that diminishes the heat conduction from the wall to the interface. Although there are not many experiments on condensation, and our modeling can reasonably explain the asymmetry of interfacial temperature discontinuity in evaporation and condensation.

Our work also clarifies a few often used mis-concepts in heat transfer analysis. Equation (10) emphasizes that the difference of the wall heat flux and latent heat equals to vapor-phase heat flux, while most phase-change heat transfer analysis assumes the wall heat flux equals the latent heat flux. Our treatment enables easy coupling to convection in the vapor phase, using well-known single phase convection results. Similarly, the liquid phase heat conduction term in Eq.(10) can also be replaced with single-phase convection results. The coupled analysis of both the liquid and the vapor phases shows that when the liquid is very thin, most of the applied temperature difference between the solid wall and the vapor phase happens between the liquid-vapor interface, leading to saturation of the evaporation and the condensation rates. Most of past work in phase-change analysis has neglected the interfacial temperature drop, leading to an overprediction of the heat and mass transfer rate. These results have important implications for modeling phase-change heat transfer processes.

**Acknowledgments**

This work is supported by MIT. The author is grateful for critical reading and comments of the manuscript by Mr. Carlos Daniel Diaz Marin, Mr. Simo Pajovic, and Dr. Lenan Zhang, Professors Hadi Ghasemi and Yangying Zhu.